\documentstyle[aps]{revtex}
\begin{document}
\title{Integrated random processes exhibiting long tails, finite moments
and 1/f spectra}
\author{Jaume Masoliver\cite{email}, Miquel Montero}
\address{Departament de F\'{\i}sica Fonamental. Universitat de
Barcelona.\\Diagonal 647, E-08028 Barcelona. Spain.}
\author{Alan McKane}
\address{Department of Theoretical Physics. University of Manchester.\\
Manchester M13 9PL, UK.}
\date{\today}
\maketitle

\begin{abstract}
A dynamical model based on a continuous addition of colored shot noises is
presented. The resulting process is colored and non-Gaussian. A general
expression for the characteristic function of the process is obtained,
which, after a scaling assumption, takes on a form that is the basis of
the results derived in the rest of the paper. One of these is an expansion
for the cumulants, which are all finite, subject to mild conditions on the
functions defining the process. This is in contrast with the L\'{e}vy
distribution --which can be obtained from our model in certain limits--
which has no finite moments. The evaluation of the power
spectrum and the form of the probability density function in the tails of
the distribution shows that the model exhibits a $1/f$ spectrum and long
tails in a natural way. A careful analysis of the characteristic function
shows that it may be separated into a part representing a L\'{e}vy
processes together with another part representing the deviation of our
model from the L\'{e}vy process. This allows our process to be viewed as a
generalization of the L\'{e}vy process which has finite moments.
\end{abstract}


\section{Introduction}

The nature of the probability distribution of stock market prices has been
discussed quantitatively for over a century~\cite{cootner}. An early
conjecture that the distribution was Gaussian was found not to be a good
fit, largely because of the long tails found in financial
data~\cite{kendall}. A later suggestion that the L\'{e}vy distribution was
a better fit seemed more promising, since this distribution does at least
have long tails~\cite{fama}. On the other hand this distribution has no
finite moments, which is a severe limitation. The solution of truncating
the distribution in order to obtain finite moments is rather {\it ad hoc}
and artificial~\cite{koponen}. Moreover, the L\'{e}vy distribution has
``too fat tails" --as opposed to the ``too thin tails" of the Gaussian--
to give a good fit to data~\cite{mantegna}. 

This paper explores in more detail a model previously
introduced~\cite{montero} in order to resolve these difficulties and give
a more complete explanation of the appearance of non-Gaussian and
self-similar fat tails in the probability distribution~\cite{mantegna},
while still keeping the important feature that all moments are finite. The
model has the L\'{e}vy distribution as a limiting form and has the form of
a type of ``Edgeworth expansion" for the L\'{e}vy
distribution. We will give the explicit form of this relationship later in
the paper (the Edgeworth series is an expansion procedure which gives
corrections to the Gaussian distribution in those cases where the Central
Limit Theorem applies~\cite{cramer}).

While the motivation for the model was the explanation of stock market
data, we want to stress in this paper the more general features of the
model which we expect will have applications in other areas including
physics. An important aspect of our integrated process is that it exhibits
a $1/f$ spectrum. We say that noise is $1/f$ if any correlated random
process has a power spectral density which is, in some range, inversely
proportional to a power of the frequency, {\it i.e.} $1/f^{\nu}$
($\nu>0$). Strictly speaking $1/f$ noise corresponds to the case when
$\nu\simeq 1$ otherwise one talks of ``flicker noise". However, here we
use the term $1/f$ noise to apply to any power law with $\nu>0$. From a
historical point of view the first experimental observation of $1/f$ noise
was made in 1925 by J. B. Johnson in studying non-stationary currents in
vacuum tubes~\cite{johnson}, and the first attempt of a theoretical
explanation was given by Schottky~\cite{schottky} in 1926. Since then
$1/f$ noise has appeared, not only in semiconductor physics~\cite{gupta},
but in many areas of physics and even natural and social
sciences~\cite{dutta}-\cite{mandelbrotb}. In spite of its ubiquity $1/f$
noise is not well understood from a theoretical point of view, and
although several mechanisms to generate the noise have been
proposed~\cite{dutta}-\cite{weissman}, as is pointed out in~\cite{milotti}
they appear to be very specialized and do not address the universality of
this noise. In this sense, we hope that the approach herein will shed some
light in the understanding of $1/f$ noise as it can be considered a
physical mechanism generating such kind of noises. 

The paper is organized as follows. In Sect. II we present the main results
of the paper for clarity, since many details are quite technical. In Sect
III we describe in detail our integrated process. In Sect. IV we impose
the self-scaling property on the probability distribution and obtain
several relevant functions. Section V is devoted to moments and cumulants
and the evaluation of the power spectrum which shows the $1/f$ character
of our process. In Section VI we study the asymptotic behavior of
the distribution while in Sect. VII we present, starting form our
distribution, an Edgeworth-type series for the L\'{e}vy distribution.
Conclusions are drawn in Sect. VIII and some more technical details are in
Appendices.

\section{Main Results}

In this section we wish to state the main results of the paper without
giving the derivations or being careful to state the range of validity for
which they hold. We hope that this will give a good indication of the
scope and nature of our results; the interested reader can then fill out
this basic framework by proceeding to later sections. 

As explained in the next section, our integrated process $X(t)$ is
a continuos superposition of colored shot noises parametrized by $u$. It
takes the form
$$
X(t)=\int_{-\infty}^{\infty}\left[
\sum_{k=1}^{\infty}A_k(u)\phi\bigl(t-T_k(u); u\bigr)\right]du.
$$
There are several components: (i) The pulse shape function, $\phi(t,u)$,
(ii) the jump amplitude $A_k(u)$ of the $k$th pulse, (iii) the jump time
$T_k(u)$ of the $k$th pulse. The jump amplitudes, $A_k(u)$, are
identically distributed random variables distributed according the
probability density function (pdf) $h(x,u)$, {\it i.e.}
$h(x,u)dx=\mbox{Prob}\{x<A_k(u)<x+dx\}$. The jump
times are assumed to follow a Poisson process with parameter $\lambda(u)$.
Therefore the model is characterized by the functions $\phi(t,u)$,
$h(x,u)$ and $\lambda(u)$. There are very few restrictions on these
functions (one is the causality condition $\phi(t,u)=0$ for $t<0$) but we
will argue that it is natural to assume the scaling forms
$$
\phi(t,u)=\phi[\lambda(u)t] \qquad {\rm and} \qquad
h(x,u)=\frac{1}{\sigma(u)}h\left[\frac{x}{\sigma(u)}\right],
$$
where $\sigma(u)$ is the standard deviation of the jump amplitudes. The
model is now specified by the four functions of a single variable $\phi$,
$h$, $\sigma$ and $\lambda$.

One of the main goals of this paper is to find the probability
distribution of the process $X(t)$, this distribution will be obtained
through the characteristic function (cf)
$\tilde{p}(\omega,t)=\langle\exp[i\omega
X(t)]\rangle,$  which is the Fourier transform of the density $p(x,t)$. By
assuming the scaling forms given above the one time distribution of $X(t)$
is explicitly given by the following cf
$$
\tilde{p}(\omega,t)=\exp\left\{-bt\omega^\alpha
\int_{0}^{\infty}\frac{dz}{z^{1+\alpha}}
\left[1-\int_0^1\tilde{h}\left[z\phi\left(\frac{bt\omega^\alpha s}
{z^\alpha}\right)\right]ds\right]\right\},
$$
where $b>0$. When $\phi(x)=\theta(x)$ is the Heaviside step function then
the input shot noises are white and $X(t)$ is the L\'{e}vy process:
$$
\tilde{p}(\omega,t)=e^{-Mt\omega^\alpha},
$$
where $M=b\int_{0}^{\infty}dz[1-\tilde{h}(z)]/z^{1+\alpha}$. This provides
us with an alternative interpretation of the L\'{e}vy process since, in
our case, L\'{e}vy processes are continuos superpositions of families of
white Poissonian shot noises. 

Contrary to L\'{e}vy processes, our integrated process
can have finite moments of any order, thus cumulants are given by
$$
\langle\langle X^n(t)\rangle\rangle=
\frac{i^{-n}\tilde{h}^{(n)}(0)}{n}(bt)^{n/\alpha}
\int_0^\infty\frac{\phi^n(x)}{x^{n/\alpha}}dx,
$$
$(n=1,2,3,\cdots)$. Note that the second cumulant, ({\it i.e.}, the
variance), is proportional to $(bt)^{2/\alpha}$ and $X(t)$ 
presents anomalous diffusion behavior (see Sect. V for a detailed
discussion on limiting values and bounds for exponent $\alpha$ and other
parameters related to the asymptotic behavior of the pulse shape function
$\phi(x)$).  

We define the stationary correlation function by
$C(\tau)=\lim_{t\rightarrow\infty}\langle X(t+\tau)X(t)\rangle$. For our
process this reads
$$
C(\tau)=b\int_0^{\infty}\frac{dz}{z^{1+\alpha}}
\int_0^{\infty}\phi\bigl(bz^{-\alpha}t'\bigr)
\phi\bigl(bz^{-\alpha}(t'+\tau)\bigr)dt'.
$$
The power spectral density of the process $X(t)$, given by the Fourier
transform of the stationary correlation, is 
$$
\tilde{C}(\omega)=\frac{K}{\omega^{1+2/\alpha}},
$$
and $X(t)$ is $1/f$ noise with exponent $\nu=1+2/\alpha$. 

We also can perform the asymptotic analysis of the probability
distribution of $X(t)$ without having to specify any particular form for
$h(x)$ and $\phi(x)$ thus keeping the maximum level of generality.
Specifically we show in Sect. VI that the center of the distribution is
approached by a L\'{e}vy distribution 
$$
\tilde{p}(\omega,t)\approx e^{-L(t)
\omega^{\delta}}\qquad\omega\rightarrow\infty,
$$
where $0<\delta<2$. We refer the reader to see Sect. VI for more details
and for the behavior of the tails of the distribution which are mainly
determined by the behavior of the jump pdf $h(x)$. 

The relation to the L\'{e}vy distribution is explored in more detail in
Sect. VII where we present and alternative (and exact) expression for the
cf which decompose the distribution of the integrated process $X(t)$ into
that of L\'{e}vy plus an additional term:
\begin{eqnarray*}
\ln\tilde{p}(\omega,t)=\ln\tilde{p}_{\rm Levy}(\omega,t)-
\frac{1}{\alpha}\int_0^\infty 
\Bigl[\tilde{h}\bigl([bt/x]^{1/\alpha}\omega\bigr)
&-&
\tilde{h}\bigl([bt/x]^{1/\alpha}\omega\phi(x)\bigr)\Bigr]dx
\nonumber \\
&+&
\int_0^\infty 
x\frac{\phi'(x)}{\phi(x)}
\biggl[1-\tilde{h}\bigl([bt/x]^{1/\alpha}\omega\phi(x)\bigr)\biggr]dx.
\end{eqnarray*}
Note that when $\phi(x)$ is the rectangular step function this equation
reduces to the L\'{e}vy distribution. Therefore, when $\phi(x)$ is a
step-like function close to the Heaviside function, this alternative
expression can be used as the starting point of an Edgeworth-type
expansion procedure giving corrections
to the L\'{e}vy distribution. 

\section{The integrated process}

Let $X(t)$ be a random process formed by a continuos superposition of
independent shot-noise processes:
\begin{equation}
X(t)=\int_{-\infty}^{\infty}Y(u,t)du,
\label{1}
\end{equation}
where for any fixed time $t$, $Y(u,t)$ are independent random variables
for different values of parameter $u$ (see Eq.~(\ref{8}) below) and for
any fixed value of $u$, $Y(u,t)$ is a colored shot-noise process
represented by a countable superposition of pulses of identical shape:
\begin{equation}
Y(u,t)=\sum_{k=1}^{\infty}A_k(u)\phi\left[t-T_k(u); u\right],
\label{2}
\end{equation}
where $T_k(u)$ marks the onset of the $k$th pulse, and $A_k(u)$ is its
amplitude. Both $T_k(u)$ and $A_k(u)$ are independent and identically
distributed random variables with probability density functions 
given by $h(a,u)$ and $\psi(t,u)$, respectively. The pulse shape
$\phi(t,u)$ has to fulfill the ``causality condition", {\it i.e.},
$\phi(t,u)=0$ for $t<0$~\cite{rice}. 

We assume that the occurrence of jumps is a Poisson process, in this case
the shot-noise $Y(t,u)$ is Markovian, and the pdf for the time interval
between jumps, $\psi(t,u)dt=\mbox{P}\{t<T_k(u)-T_{k-1}(u)<t+dt\}$, is
exponential:
\begin{equation}
\psi(t,u)=\lambda(u)e^{-\lambda(u)t}\qquad(t\geq 0),
\label{3}
\end{equation}
where $\lambda(u)$ is the mean jump frequency, {\it i.e.}, $1/\lambda(u)$
is the mean time between two consecutive jumps~\cite{maso87}. We recall
that jump amplitudes $A_k(u)$ are identically distributed (for all
$k=1,2,3,\cdots$) and independent random variables (for all $k$ and $u$).
In what follows we will assume that they have zero mean and a pdf,
$h(x,u)dx=\mbox{P}\{x<A_k(u)<x+dx\}$, depending on a single ``dimensional"
parameter which, without loss of generality, we assume to be the standard
deviation of jumps $\sigma(u)=\sqrt{\langle A_k^2(u)\rangle}$. That is,
\begin{equation}
h(x,u)=\frac{1}{\sigma(u)}h\left[\frac{x}{\sigma(u)}\right].
\label{4}
\end{equation}

Before proceeding further with the probability distribution of the
integrated process $X(t)$ given by Eq.~(\ref{1}), we note that
following Rice's method~\cite{rice} one can easily obtain all the
probability distributions of the shot noise $Y(t,u)$ via their
cf's 
\begin{equation}
\tilde{p}_Y(\omega_1,t_1;\cdots;\omega_n,t_n;u)
=\left\langle\exp\left[i\sum_{k=1}^{n}\omega_k Y(t_k,u)\right]
\right\rangle.
\label{cf}
\end{equation}
In Appendix A we show that
\begin{equation}
\ln\tilde{p}_Y(\omega,t;u)=-\lambda(u)\left[t-
\int_0^t\tilde{h}[\omega\sigma(u)\phi(t',u)]dt'\right],
\label{5}
\end{equation}
and (supposing that $t_2>t_1$)
\begin{eqnarray}
\ln\tilde{p}_Y(\omega_1,t_1;\omega_2,t_2;u)=-\lambda(u)\Biggl[t_2-
\int_0^{t_1}\tilde{h}[\omega_1\sigma(u)\phi(t',u)&+&
\omega_2\sigma(u)\phi(t'+t_2-t_1,u)]dt'\nonumber\\ &-&
\int_0^{t_2-t_1}\tilde{h}[\omega_2\sigma(u)\phi(t',u)]dt'\Biggr],
\label{5a}
\end{eqnarray}
where $\tilde{h}(\omega)$ is the Fourier transform of the jump pdf $h(x)$. 

Let us now evaluate the probability distribution of the integrated process
$X(t)$. In terms of the cumulants, $\langle\langle Y(t,u)\rangle\rangle$
of the shot noise $Y(t,u)$ we see that the one time characteristic
function of $X(t)$ can be written as
\begin{equation}
\tilde{p}_X(\omega,t)=\exp\left\{\sum_{k=1}^{\infty}\frac{(i\omega)^k}{k!}
\left\langle\left\langle\left[\int_{-\infty}^{\infty}Y(u,t)du\right]^k
\right\rangle\right\rangle\right\}.
\label{6}
\end{equation}
That is,
\begin{equation}
\ln\tilde{p}_X(\omega,t)=\sum_{k=1}^{\infty}\frac{(i\omega)^k}{k!}
\int_{-\infty}^{\infty}\cdots\int_{-\infty}^{\infty}
\langle\langle Y(u_1,t)\cdots Y(u_k,t)\rangle\rangle 
du_1\cdots du_k
\label{7}
\end{equation}
But, by our assumptions on the process $Y(u,t)$ we have
\begin{equation}
\langle\langle Y(u_1,t)\cdots Y(u_k,t)\rangle\rangle=
\langle\langle Y^k(u_1,t)\rangle\rangle
\delta(u_1-u_2)\cdots\delta(u_{k-1}-u_k).
\label{8}
\end{equation}
Therefore,
$$
\ln\tilde{p}_X(\omega,t)=
\sum_{k=1}^{\infty}\frac{(i\omega)^k}{k!}
\int_{-\infty}^{\infty}\langle\langle Y^k(u,t)\rangle\rangle du,
$$
that is,
\begin{equation}
\ln\tilde{p}_X(\omega,t)=
\int_{-\infty}^{\infty}\ln\tilde{p}_Y(\omega,t;u)du.
\label{9}
\end{equation}
Note that this line of reasoning can be easily extend to the $n$th time
distribution, with the result:
\begin{equation}
\ln\tilde{p}_X(\omega_1,t_1;\cdots;\omega_n,t_n)=
\int_{-\infty}^{\infty}
\ln\tilde{p}_Y(\omega_1,t_1;\cdots;\omega_n,t_n;u)du.
\label{10}
\end{equation}

Going back to our integrated process we have from
Eqs.~(\ref{5})-(\ref{5a}) and~(\ref{9})-(\ref{10}) that the one
time characteristic function of $X(t)$ reads
\begin{equation}
\ln\tilde{p}(\omega,t)=-\int_{-\infty}^{\infty}du\lambda(u)
\left[t-\int_0^t\tilde{h}[\omega\sigma(u)\phi(t',u)]dt'\right],
\label{11}
\end{equation}
while the two time cf is ($t_2>t_1$)
\begin{eqnarray}
\ln\tilde{p}(\omega_1,t_1;\omega_2,t_2)=
\int_{-\infty}^{\infty}du\lambda(u)
\Biggl\{\int_0^{t_1}dt'\biggl[\tilde{h}[\omega_1\sigma(u)\phi(t',u)&+&
\omega_2\sigma(u)\phi(t'+t_2-t_1,u)]-1\biggr]\nonumber\\
&+&
\int_0^{t_2-t_1}dt'\biggl[\tilde{h}[\omega_2\sigma(u)\phi(t',u)]-1\biggr]
\Biggr\},
\label{11a}
\end{eqnarray}
where we have dropped the subscript $X$. Obviously these are formal
expressions, as long as we do not provide the functional dependence of
$\lambda(u)$ and $\sigma(u)$ on the parameter $u$. We will do so in the
next section using scaling arguments. 

\section{Scaling}

In order to proceed further we need to specify the functional form of 
$\lambda(u)$ and $\sigma(u)$. Of course that form will depend on the
specific features of the problem at hand. At this point we
choose what seems to us one of the most general ways of proceeding, we
thus suppose that our integrated process $X(t)$ possesses self-scaling
properties. Following this path we must first assume that the pulse
function is of the form
\begin{equation}
\phi(u,t)=\phi[\lambda(u)t],
\label{12}
\end{equation}
which turns $\phi(u,t)$ into a function of the single dimensionless
variable $\lambda(u)t$. Substituting this into Eq.~(\ref{11}), defining
new integration variables $s=t'/t$ and $z=\omega\sigma(u)$, and supposing
that $\sigma(-\infty)=0$ and $\sigma(\infty)=\infty$, we obtain
\begin{equation}
\ln\tilde{p}(\omega,t)=-\int_{0}^{\infty}dz
\frac{\lambda t}{\omega\sigma'}
\left\{1-\int_0^1\tilde{h}[z\phi(\lambda st)]ds\right\},
\label{13}
\end{equation}
where the prime on $\sigma$ denotes derivative. We now impose the
self-scaling property on the cf, that is, we assume that
$\tilde{p}(\omega,t)$ is a function of the single variable $\omega
t^{1/\alpha}$:
\begin{equation}
\tilde{p}(\omega,t)=f(\omega t^{1/\alpha}).
\label{14}
\end{equation}
On the other hand we note that in Eq.~(\ref{13}), the quantities $\sigma'$
and $\lambda$ are functions of $z$ and $\omega$. Then scaling~(\ref{14})
implies 
\begin{equation}
\lambda=B(z)\omega^\alpha,\qquad
\frac{\lambda}{\omega\sigma'}=A(z)\omega^\alpha,
\label{15}
\end{equation}
where $A(z)$ and $B(z)$ are arbitrary functions to be determined. From
these two relations we get $\sigma'=C(z)/\omega$, where $C(z)=B(z)/A(z)$.
In the original variable $u$ we have
\begin{equation}
\sigma'=\frac{C(\omega\sigma(u))}{\omega}.
\label{16}
\end{equation}
But $\sigma'=\sigma'(u)$ is independent of $\omega$. Therefore, the
unknown function $C$ has to be of the form $C(\omega\sigma)=k\omega\sigma$
where $k$ is a constant. Hence $\sigma'(u)=k\sigma(u)$, whence
$\sigma(u)=\sigma_0e^{ku}$. Finally absorbing the constant $k$ inside the
variable $u$ we obtain the functional dependence of the jump variance
$\sigma$ on parameter $u$:
\begin{equation}
\sigma(u)=\sigma_0e^u,
\label{17}
\end{equation}
where $\sigma_0$ is a constant. Moreover we see from Eq.~(\ref{15}) that
$\lambda=B(\omega\sigma(u))\omega^\alpha$. But, again $\lambda=\lambda(u)$
is independent of $\omega$. In consequence 
$B(\omega\sigma)=b\ (\omega\sigma)^{-\alpha}$ where $b$ is
an arbitrary  constant. Substituting this into the first relation of
Eq.~(\ref{15}) yields the ``dispersion relation" between the mean
frequency $\lambda$ and the jump variance $\sigma$:
\begin{equation}
\lambda=b/\sigma^\alpha.
\label{18}
\end{equation}
The functional dependence of $\lambda$ on $u$ is obtained by combining
Eqs.~(\ref{17})-(\ref{18}),
\begin{equation}
\lambda(u)=\lambda_0e^{-\alpha u},
\label{19}
\end{equation}
where $\lambda_0=b/\sigma_0^\alpha$, or equivalently
\begin{equation}
b=\lambda_0\sigma_0^\alpha.
\label{20}
\end{equation}
Collecting results we see from Eq.~(\ref{13}) and
Eqs.~(\ref{17})-(\ref{20}) that the one-time characteristic function
reads 
\begin{equation}
\tilde{p}(\omega,t)=\exp\left\{-bt\omega^\alpha
\int_{0}^{\infty}\frac{dz}{z^{1+\alpha}}
\left[1-\int_0^1\tilde{h}\left[z\phi\left(bt\omega^\alpha s/z^\alpha
\right)\right]ds\right]\right\}.
\label{21}
\end{equation}
It is sometimes convenient to rewrite this equation and use the
alternative form of $\tilde{p}(\omega,t)$ given by 
\begin{equation}
\tilde{p}(\omega,t)=\exp\left\{-bt\int_{0}^{\infty}\frac{dz}{z^{1+\alpha}}
\left[1-\int_0^1\tilde{h}\left[z\omega\phi\left(bts/z^\alpha\right)
\right]ds\right]\right\},
\label{21a}
\end{equation}
or equivalently
\begin{equation}
\tilde{p}(\omega,t)=\exp\left\{b\int_{0}^{\infty}\frac{dz}{z^{1+\alpha}}
\int_0^t dt'\left[\tilde{h}\left[z\omega\phi\left(bt'/z^\alpha\right)
\right]-1\right]\right\}.
\label{21b}
\end{equation}

Starting from Eq.~(\ref{11a}) and following an analogous reasoning based
on the scaling assumption, we obtain the following expression for the two
time characteristic function of the integrated process (with $t_2>t_1$)
\begin{eqnarray}
\ln\tilde{p}(\omega_1,t_1;\omega_2,t_2)=
b\int_{0}^{\infty}\frac{dz}{z^{1+\alpha}}
\Biggl\{\int_0^{t_1}dt'\biggl[
\tilde{h}[z\omega_1\phi(bz^{-\alpha}t')&+&
z\omega_2\phi(bz^{-\alpha}(t'+t_2-t_1))]-1\biggr] \nonumber \\ 
&+&\int_0^{t_2-t_1}dt'\biggl[\tilde{h}[z\omega_2\phi(bz^{-\alpha}t')]-1
\biggr]\Biggr\}.
\label{21c}
\end{eqnarray}
Equations~(\ref{21})-(\ref{21c}) are some of the key results of the
paper. Since, as we will see next, they constitute a generalization of the
L\'{e}vy distribution with finite moments.

\section{Moments, Cumulants and Power Spectrum}

We first note from Eq.~(\ref{21}) that if the pulse shape function
is the Heaviside step function:
\begin{equation}
\phi(t)=\cases{1,&if $t>0$\cr
                 0, &otherwise,\cr}
\label{theta}
\end{equation}
then the integrated process $X(t)$ is identically a L\'{e}vy process,
regardless the jump pdf $h(x)$:
\begin{equation}
\tilde{p}(\omega,t)=\exp\left[-Mt\omega^\alpha\right],
\label{22}
\end{equation}
where 
\begin{equation}
M=b\int_{0}^{\infty}\frac{dz}{z^{1+\alpha}}
\left[1-\tilde{h}(z)\right].
\label{23}
\end{equation}
Therefore, following our model, L\'{e}vy processes can be viewed as a
continuos superposition of families of rectangular pulses occurring at
random Poisson times. The usual range
of the exponent $\alpha$ in L\'{e}vy flights is $0<\alpha<2$. In such a
case  $X(t)$ has no finite moment but the first one~\cite{shlesinger}. In
actual situations, one is unlikely to meet with perfect rectangular
pulses (showing sudden changes) in such a case all moments can be finite
and are easily evaluated from Eq.~(\ref{21a}) through
$$
\langle X^n(t)\rangle=\left.i^{-n}\frac{\partial^n\tilde{p}(\omega,t)}
{\partial\omega^n}\right|_{\omega=0}.
$$
Thus for instance the second moment is given by (note that due to
Eq.~(\ref{4}) $\tilde{h}''(0)=-1$)
\begin{equation}
\langle X^2(t)\rangle=bt\int_0^\infty\frac{dz}{z^{\alpha-1}}
\int_0^1\phi^2(bts/z^\alpha)ds.
\label{24a}
\end{equation}
As an illustrative example suppose that our pulse function has the form
\begin{equation}
\phi(t)=\cases{1-e^{-kt}, &if $t>0$\cr
                 0, &otherwise,\cr}
\label{24b}
\end{equation}
where $k>0$ is a constant (note that if $k$ is large then $\phi(t)$
approaches to the rectangular pulse~(\ref{theta})). In the Appendix B we
show that (see also Eq.~(\ref{cumulant3}) below)
\begin{equation}
\langle X^2(t)\rangle=D t^{2/\alpha},
\qquad (2>\alpha>2/3),
\label{25}
\end{equation}
where $D=\alpha b^2k^{-1+2/\alpha}(1-2^{2/\alpha-2})\Gamma(2-2/\alpha)/
(2-\alpha)$. Since $1<(2/\alpha)<3$, Eq.~(\ref{25}) clearly shows a
superdiffusive behavior. 

In fact we can easily obtain a closed expression not for the $n$th moment
but for the $n$th cumulant
$$
\langle\langle X^n(t)\rangle\rangle\equiv
i^{-n}\left.\frac{\partial^n\ln\tilde{p}(\omega,t)}
{\partial\omega^n}\right|_{\omega=0}.
$$
>From Eq.~(\ref{21a}) we have
$$
\langle\langle X^n(t)\rangle\rangle=i^{-n}bt\tilde{h}^{(n)}(0)
\int_0^\infty z^{n-1-\alpha}dz\int_0^1\phi^n(bts/z^\alpha)ds.
$$
If in the double integral on the right hand side of this
equation we define a new integration variable $x$ by
$s=(z^{\alpha}/bt)x$ and exchange the order of
integration we get
$$
\int_0^\infty z^{n-1-\alpha}dz\int_0^1\phi^n(bts/z^\alpha)ds=
\frac{1}{bt}\int_0^\infty\phi^n(x)dx
\int_0^{(bt/x)^{1/\alpha}}z^{n-1}dz,
$$
but the last integral is trivially evaluated, and for the $n$th cumulant
we have
\begin{equation}
\langle\langle X^n(t)\rangle\rangle=
\frac{i^{-n}\tilde{h}^{(n)}(0)}{n}(bt)^{n/\alpha}
\int_0^\infty\frac{\phi^n(x)}{x^{n/\alpha}}dx.
\label{cumulant}
\end{equation}
Taking into account that
$\tilde{h}^{(n)}(0)=0$ for $n$ odd (we have assumed a symmetric jump
distribution $h(x)$) we write
\begin{equation}
\langle\langle X^{2n-1}(t)\rangle\rangle=0,
\label{cumulant2a}
\end{equation}
and
\begin{equation}
\langle\langle X^{2n}(t)\rangle\rangle=
\frac{(-1)^{n}\tilde{h}^{(2n)}(0)}{2n}(bt)^{2n/\alpha}
\int_0^\infty\frac{\phi^{2n}(x)}{x^{2n/\alpha}}dx,
\label{cumulant2}
\end{equation}
($n=1,2,3,\cdots$). In order to check the convergence of these expressions
and the existence of moments, we first assume that
\begin{equation}
\phi(x)\sim x^\beta,\qquad(x\rightarrow 0),
\label{beta}
\end{equation}
($\beta>0$) then the convergence of the integral on the right hand side of
Eq.~(\ref{cumulant2}) as $x\rightarrow 0$ implies that the
scaling exponent $\alpha$ has a lower bound:
\begin{equation}
\alpha>\frac{2n}{1+2n\beta}.
\label{*}
\end{equation}
On the other hand if we assume that 
\begin{equation}
\phi(x)\sim x^\gamma,\qquad(x\rightarrow\infty),
\label{gamma}
\end{equation}
then the convergence of~(\ref{cumulant2}) when
$x\rightarrow\infty$ implies that the scaling exponent $\alpha$ also has
an upper bound:
\begin{equation}
\frac{1}{\alpha}>\gamma+\frac{1}{2n}.
\label{bound0}
\end{equation}
Moreover when $\gamma\geq 0$ then if
\begin{equation}
\frac{1}{\gamma+1/2}>\alpha>\frac{1}{\beta}
\label{bounda}
\end{equation}
all cumulants will exist. Note that Eq.~(\ref{bound0}) holds whenever
$\gamma\geq-1/2n$. Therefore, $\gamma\geq 0$ is a sufficient condition for
its validity. On the other hand if $\gamma<-1/2$ there is no upper bound
on the accepted values of $\alpha$. We also observe that for a step-like
function, as that of Eq.~(\ref{24b}), where $\gamma=0$ then all cumulants
will exist if 
$$
2>\alpha>1/\beta.
$$
Finally, for any integrable function $\phi(t)$ over $[0,\infty)$ there is
no upper bound for $\alpha$ and the only condition on $\alpha$ for having
all moments finite is that $\alpha>1/\beta$.
 
We close this discussion on moments and cumulants with an example. Suppose
that the pulse function is given by the step-like function~(\ref{24b}). In
this case $\gamma=0$, $\beta=1$ and all moments (and cumulants) will exist
if $1<\alpha<2$. Cumulants are given by
Eqs.~(\ref{cumulant2a})-(\ref{cumulant2}). In Appendix B we show that
\begin{equation}
\int_0^{\infty}\frac{(1-e^{-kx})^n}{x^n}dx=
A_nk^{-1+n/\alpha}\Gamma(1+n-n/\alpha),
\label{integral}
\end{equation}
where the numbers $A_n$ are given by Eq.~(\ref{b3}) of Appendix B.
Finally,
\begin{equation}
\langle\langle X^{2n}(t)\rangle\rangle=
\frac{(-1)^{n}\tilde{h}^{(2n)}(0)}{2n}A_{2n}
k^{-1+2n/\alpha}\Gamma(1+2n-2n/\alpha)(bt)^{2n/\alpha}.
\label{cumulant3}
\end{equation}

We finish this section evaluating the power spectrum of the integrated
process $X(t)$. Let us first evaluate the correlation function
$$
\langle X(t+\tau)X(t)\rangle=-\left.
\frac{\partial^2}{\partial\omega_1\omega_2}
\tilde{p}(\omega_1,t;\omega_2,t+\tau)\right|_{\omega_1=\omega_2=0}.
$$
>From Eq.~(\ref{21c}) we get
\begin{equation}
\langle X(t+\tau)X(t)\rangle=
b\int_0^{\infty}\frac{dz}{z^{1+\alpha}}\int_0^t\phi(bz^{-\alpha}t')
\phi(bz^{-\alpha}(t'+\tau)).
\label{correlation1}
\end{equation}
Let $C(\tau)$ be the correlation function in the stationary limit
$t\rightarrow\infty$, {\it i.e.}
$$
C(\tau)=\lim_{t\rightarrow\infty}\langle X(t+\tau)X(t)\rangle.
$$
>From Eq.~(\ref{correlation1}) we have
\begin{equation}
C(\tau)=b\int_0^{\infty}\frac{dz}{z^{1+\alpha}}
\int_0^{\infty}\phi(bz^{-\alpha}t')
\phi(bz^{-\alpha}(t'+\tau)).
\label{correlation2}
\end{equation}
Note that the (stationary) variance $C(0)=\infty$ which agrees with the
superdiffusive behavior of $X(t)$ given by Eq.~(\ref{25}). 

The power spectral density of our process is thus given by the Fourier
transform of the stationary correlation function
$$
\tilde{C}(\omega)=\int_{-\infty}^{\infty}e^{-i\omega\tau}C(\tau)d\tau.
$$
Substituting Eq.~(\ref{correlation2}) into this equation, performing
simple changes of variables and taking into account the causality of the
pulse function $\phi(t)$ we finally obtain
\begin{equation}
\tilde{C}(\omega)=\frac{K}{\omega^{1+2/\alpha}},
\label{spectrum1}
\end{equation}
where
\begin{equation}
K=\frac{b^{2/\alpha}}{\alpha}
\int_0^{\infty}\xi^{2/\alpha}|\tilde\phi(\xi)|^2d\xi,
\label{K}
\end{equation}
and $\tilde\phi(\xi)$ is the Fourier transform of $\phi(t)$. The power
spectral density $\tilde{C}(\omega)$ exists if the integral 
$$
J=\int_0^{\infty}\xi^{2/\alpha}|\tilde\phi(\xi)|^2d\xi<\infty.
$$
In order to prove the existence of $J$ we first need that the Fourier
transform of the pulse function, $\tilde{\phi}(\xi)$, exists. Note that
any step or step-like function does not have a Fourier transform and
consequently the power spectrum is infinite. For the existence of
$\tilde{\phi}(\xi)$ it suffices that $\phi(x)$ be absolutely integrable,
and from the asymptotic behavior given by Eq.~(\ref{gamma}):
$$
\phi(x)\sim x^{\gamma}\qquad(x\rightarrow\infty),
$$
we have to impose that $\gamma<-1$. This in turn implies that
$\tilde{\phi}(\xi)\sim\xi^{-1-\gamma}$ as $\xi\rightarrow 0$ and, since
$\gamma<-1$, the integral $J$ at its lower limit is always finite for any
$\alpha>0$. On the other hand if $\phi(x)$ satisfies Eq.~(\ref{beta}) as
$x\rightarrow 0$, then $\tilde{\phi}(\xi)\sim\xi^{-1-\beta}$
as $\xi\rightarrow\infty$. Hence, $J$ will be finite if
$\alpha>1/(\beta+1/2)$ (see also Eq.~(\ref{*})). Therefore, the
process $X(t)$ has a finite power spectrum if $\phi(x)$ is absolutely
integrable on the real line, and $X(t)$ has a finite second cumulant,
Eq.~(\ref{*}), {\it i.e.}
\begin{equation}
\alpha>\frac{1}{\beta+1/2}.
\label{boundc}
\end{equation}
In such a case, we see from Eq.~(\ref{spectrum1}) that $X(t)$ is
$1/f$ noise with exponent $\nu=1+2/\alpha$. Moreover (recall that
$\alpha>0$, and Eq.~(\ref{boundc}))
\begin{equation}
1<\nu<2(1+\beta).
\label{boundf}
\end{equation}
Finally, when $\phi(x)$ is analytic at $x=0$ then $\beta=1,2,3,\cdots$ is
a positive integer and $X(t)$ is flicker noise $1/f^{\nu}$ with
$1<\nu<2(1+n)$ ($n=1,2,3,\cdots$).

\section{Asymptotic Behavior}

We will now examine the asymptotic behavior of the one time probability
density function (pdf) of the integrated process $X(t)$,
$p(x,t)dx=\mbox{P}\{x<X(t)<x+dx\}$. For this analysis we distinguish two
regions: the ``center" ($x\rightarrow 0$) and the ``tails"
($x\rightarrow\pm\infty$) of the distribution. We cannot have a
closed expression for the pdf $p(x,t)$ until the pulse function $\phi(x)$
and the jump pdf $h(x)$ are both specified. Therefore, we will perform the
asymptotics on the cf $\tilde{p}(\omega,t)$. As a well known feature of
the harmonic analysis the center of the distribution is determined by the
large $\omega$ behavior of the cf, while the tails are determined by
$\tilde{p}(\omega,t)$ when $\omega\rightarrow 0$~\cite{weissb}. 

We deal first with the center of the distribution where 
$\omega\rightarrow\infty$. If we assume that 
the pulse function $\phi(x)$, as $x\rightarrow\infty$, satisfies
Eq.~(\ref{gamma}):
$$
\phi(x)\sim x^{\gamma},\qquad(x\rightarrow\infty),
$$
then
$$
\tilde{h}\left[z\phi\left(bt\omega^\alpha s/z^\alpha\right)\right]
\sim\tilde{h}\left[\left(bt\omega^\alpha\right)^\gamma
z^{1-\alpha\gamma}s^{\gamma}\right], \qquad(\omega\rightarrow\infty).
$$
Substituting this into Eq.~(\ref{21}) and performing the change of
variables $\xi=(bt\omega^\alpha)^\gamma z^{1-\alpha\gamma}$ we obtain the
following L\'{e}vy distribution:
\begin{equation}
\tilde{p}(\omega,t)\simeq
\exp\left\{-L(bt)^{1/(1-\alpha\gamma)}
\omega^{\alpha/(1-\alpha\gamma)}\right\},\qquad(\omega\rightarrow\infty),
\label{26a}
\end{equation}
where 
\begin{equation}
L=\frac{1}{1-\alpha\gamma}\int_{0}^{\infty}
\frac{d\xi}{\xi^{(1+\alpha-\alpha\gamma)/(1-\alpha\gamma)}}
\left[1-\int_0^1\tilde{h}(\xi s^\gamma)ds\right].
\label{L}
\end{equation}
Note that due to the bounds discussed above (see Eq.~(\ref{bound0})) we
have $1-\alpha\gamma>\alpha/2$ hence the L\'{e}vy exponent in
Eq.~(\ref{26a})
satisfies
\begin{equation}
0<\frac{\alpha}{1-\alpha\gamma}<2,
\label{exponent}
\end{equation}
and Eq.~(\ref{26a}) is well defined. We also note that 
for a step-like pulse function $\phi(t)$ where
$\gamma=0$ we obtain the same L\'{e}vy distribution, Eq.~(\ref{22}),
that satisfies the model for sudden pulses~(\ref{theta}). Therefore, for
any pulse shape function satisfying condition~(\ref{gamma}) the center of
the pdf is given by a L\'{e}vy distribution. 

Let us now obtain an asymptotic expression of the pdf $p(x,t)$ when
$x\rightarrow\pm\infty$, which will be valid if $\phi(t)$ obeys
Eqs.~(\ref{beta})
and~(\ref{gamma}), and the exponent $\alpha$ is bounded by 
$$
\beta^{-1}<\alpha<1/(\gamma+1/2).
$$
In this case, we see from Eq.~(\ref{bounda}) that all cumulants exist. So
taking the $\omega\rightarrow 0$ limit of Eq~(\ref{21a}) we find
\begin{equation}
\tilde{p}(\omega,t)\sim 1-bt
\int_{0}^{\infty}\frac{dz}{z^{1+\alpha}}
\left[1-\int_0^1
\tilde{h}\left[z\omega\phi\left(bts/z^\alpha\right)\right]dz\right].
\label{27a}
\end{equation}
The Fourier inversion of Eq.~(\ref{27a}) yields
\begin{equation}
p(x,t)\sim bt
\int_{0}^{\infty}\frac{dz}{z^{1+\alpha}}
\int_0^1h(x,z,s)dz\qquad(x\rightarrow\pm\infty),
\label{28a}
\end{equation}
where we have dropped delta function terms which have no contribution as
$x\rightarrow\infty$. Moreover assuming symmetric jump distributions
$h(x)$ we have
$$
h(x,z,s)\equiv\frac{1}{\pi}\int_0^\infty
\tilde{h}\left[z\omega\phi\left(bts/z^\alpha\right)\right]
\cos\omega x\ d\omega=
\frac{1}{z\phi(bts/z^\alpha)}h\left[\frac{x}{z\phi(bts/z^\alpha)}\right].
$$
Substituting this into Eq.~(\ref{28a}) we see that  
\begin{equation}
p(x,t)\sim bt
\int_{0}^{\infty}\frac{dz}{z^{2+\alpha}}
\int_0^1\frac{ds}{\phi(bts/z^\alpha)}
h\left[\frac{x}{z\phi(bts/z^\alpha)}\right]
\qquad(x\rightarrow\pm\infty).
\label{29a}
\end{equation}
Therefore the tails of the distribution are determined by the jump pdf
$h(x)$ and the pulse shape function $\phi(t)$.

Finally for the rectangular pulse~(\ref{theta}) we have
\begin{equation}
p(x,t)\sim\frac{bt}{|x|^{1+\alpha}}
\int_{0}^{\infty}y^{\alpha}h(y)dy,
\label{30a}
\end{equation}
which agrees with the expected tail behavior of the L\'{e}vy
distribution~\cite{shlesinger}.

\section{Relation to the L\'{e}vy distribution}

In section V we obtained general
expressions~(\ref{cumulant2a})-(\ref{cumulant}) for all of the
cumulants, from which it follows that
\begin{eqnarray} 
\ln\tilde{p}(\omega,t)&=&\sum^{\infty}_{n=1}\frac{i^{n}\omega^{n}}{n!}
\langle\langle X^{n}(t)\rangle\rangle\nonumber\\ 
&=&\sum^{\infty}_{n=1}\frac{(-1)^n\omega^{2n}\tilde{h}^{(2n)}(0)}
{2n(2n)!}(bt)^{2n/\alpha}\int^{1}_{0}\frac{\phi^{2n}(x)}{x^{2n/\alpha}}dx.
\label{6-1}
\end{eqnarray}
In addition, in section VI, we showed that $p(x,t)$ was a L\'{e}vy
distribution
at the center of the distribution ($x\rightarrow 0$) and took the
form~(\ref{29a}) in the tails ($x\rightarrow\pm\infty$) of the
distribution. In this section we will show how the distribution can be
separated into a L\'{e}vy distribution plus an additional term. This term
takes the form of a single integral which can be evaluated once the
functions $\phi$ and $h$ have been specified. 

We begin the analysis by changing variables from $s$ to 
$x=bts/z^{\alpha}$ ($z$ fixed) in Eq.~(\ref{21a}). This gives
\begin{eqnarray}
\ln\tilde{p}(\omega,t)&=&-\int^{\infty}_{0}\frac{dz}{z}
\int^{bt/z^{\alpha}}_{0}dx\left\{1-
\tilde{h}\left[z\omega\phi(x)\right]\right\}\nonumber\\
&=&-\int^{\infty}_{0}dx\int^{(bt/x)^{1/\alpha}}_{0}
\frac{dz}{z}\left\{1 -
\tilde{h}\left[z\omega\phi(x)\right]\right\}
\label{6-2}
\end{eqnarray}
changing the order of integration. At this point we factor out the
contribution from the L\'{e}vy process by writing~(\ref{6-2}) as
$$
-\int^{\infty}_{0}dx\int^{(bt/x)^{1/\alpha}}_{0}
\frac{dz}{z}\left\{1-\tilde{h}\left[z\omega\right]\right\}
-\int^{\infty}_{0}dx\int^{(bt/x)^{1/\alpha}}_{0}
\frac{dz}{z}\left\{\tilde{h}\left[z\omega\right] 
-\tilde{h}\left[z\omega\phi(x)\right]\right\},
$$
or, after defining
\begin{equation}
\tilde{g}(\omega,t)\equiv1-\tilde{h}(\omega,t),
\label{h}
\end{equation}
as
$$
-\int^{\infty}_{0}dx\int^{(bt/x)^{1/\alpha}}_{0}
\frac{\tilde{g}(\omega,t)}{z}dz
+\int^{\infty}_{0}dx\int^{(bt/x)^{1/\alpha}}_{0}
\frac{dz}{z}\left\{\tilde{g}\left[z\omega\right] 
-\tilde{g}\left[z\omega\phi(x)\right]\right\},
$$
The first term is just $\ln\tilde{p}(\omega,t)$ for the L\'{e}vy processes 
(see Eqs.~(\ref{22})-(\ref{23})). The second term can be simplified by
first writing it as 
$$
\int_0^\infty
dx\int^{(bt/x)^{1/\alpha}}_{(bt/x)^{1/\alpha}\phi}
\frac{\tilde{g}(z\omega)}{z}dz
$$
and then integrating by parts to give
\begin{equation}
\left|x\int^{(bt/x)^{1/\alpha}}_{(bt/x)^{1/\alpha}\phi}
\frac{\tilde{g}(z\omega)}{z}dz\right|_{x=0}^{x=\infty}-
\int_0^\infty dx x\frac{\partial}{\partial x}
\int^{(bt/x)^{1/\alpha}}_{(bt/x)^{1/\alpha}\phi}
\frac{\tilde{g}(z\omega)}{z}dz.
\label{parts}
\end{equation}
We assume that $\tilde{h}(\omega)$ is analytic at
$\omega=0$ and integrable, then
$$
\tilde{g}(\omega)\sim\omega^2\quad(\omega\rightarrow 0)\qquad{\rm
and}\qquad \tilde{g}(\omega)\rightarrow 1\quad(\omega\rightarrow\infty),
$$
and since $0<\alpha<2$ the first term in Eq.~(\ref{parts}) is zero.
Finally
\begin{eqnarray}
\ln\tilde{p}(\omega,t)=\ln\tilde{p}_{\rm Levy}(\omega,t)+
\frac{1}{\alpha}\int_0^\infty
\Bigl[\tilde{g}\bigl([bt/x]^{1/\alpha}\omega\bigr)
&-&
\tilde{g}\bigl([bt/x]^{1/\alpha}\omega\phi(x)\bigr)\Bigr]dx
\nonumber \\
&+&
\int_0^\infty x\frac{\phi'(x)}{\phi(x)}
\tilde{g}\bigl([bt/x]^{1/\alpha}\omega\phi(x)\bigr)dx,
\label{6-4}
\end{eqnarray}
where $\phi'(x)$ is the derivative of the pulse shape function and
$$
\ln\tilde{p}_{\rm Levy}(\omega,t)=-Mt\omega^\alpha
$$
where $M$ is given by Eq.~(\ref{23}). Note that when $\phi(x)$ is the
Heaviside step function the integrals on the right hand side of
Eq.~(\ref{6-4}) vanish and Eq.~(\ref{6-4}) reduces to the L\'{e}vy
distribution. Therefore, we can look at the second term on the right hand
side of Eq.~(\ref{6-4}) as a correction to the L\'{e}vy distribution when
$\phi(x)$ is not exactly a Heaviside function but a step-like function
very close to the Heaviside function. This may be evaluated, in principle,
for any given $\phi$ and $\tilde{h}$. For instance we could take $\phi$ to
be of the form~(\ref{24b}) with $k$ large and the Lorentzian
$$
\tilde{g}(\omega)=\frac{\omega^2/2}{1+\omega^2/2},
$$ 
corresponding to $h(x)=e^{-\sqrt{2}|x|}/\sqrt{2}$. The form of the
correction terms depends on the choice of the functions to an extent, and
so we will not discuss the explicit form it takes here.

\section{Conclusions}

In this paper we have presented and analyzed a dynamical model based on a
process which is a superposition of colored Poisson noises. The model was
shown to have several attractive features. The probability density
function has long tails which emerged in a natural way and, unlike the
L\'{e}vy distribution, all the moments of the distribution are finite. We
believe that these properties make the distribution an ideal candidate for
describing stock market prices~\cite{montero}. 

A property what may have more relevance to physics and other natural
sciences is the appearance of $1/f$ noise in the power spectrum. Once
again we would like to stress that this result flowed naturally from the
nature of the model and the scaling assumptions which reduce $h$ and
$\phi$ from functions of two variables to functions af a single variable. 

In a more mathematical context, we believe that the decomposition of the
cf of our model into that of the L\'{e}vy plus additional terms is
interesting, both as an example of an Edgeworth-type expansion and for the
nature of the corrections to the L\'{e}vy distributions when the
parameters of our model are chosen so that our distribution is near to the
L\'{e}vy one. 

There are still some open questions. One of them is the extension of the
model to the increments of the process
$Z(\tau,t-t_0)=X(t-t_0+\tau)-X(t-t_0)$ $(t>t_0)$,  since in this case we
believe that the process $Z(\tau,t-t_0)$ becomes stationary when it starts
in the infinite past ($t_0\rightarrow\infty$). Another interesting and
open question is the actual application of the model to financial time
series where some non-white correlation is observed~\cite{lo}. Both points
are presently being investigated.

\acknowledgements
This work has been supported in part by Direcci\'on General de
Investigaci\'on Cient\'{\i}fica y T\'ecnica under contract No. PB96-0188
and Project No. HB119-0104, and by Generalitat de Catalunya under contract
No. 1998 SGR-00015 (JM and MM) AM thanks the British Council for support
under ``Acciones Integradas" scheme.

\appendix

\section{Characteristic function for colored shot noise}

By generalizing Rice's method~\cite{rice}, we will now obtain the
probability distribution of the shot noise $Y(u,t)$ defined by
Eq.~(\ref{2}):
\begin{equation}
Y(u,t)=\sum_{k=1}^{\infty}A_k(u)\phi\left[t-T_k(u); u\right],
\label{a1}
\end{equation}
where we assume that the random variables $A_k(u)$ and $T_k(u)$ are
identically distributed and statistically independent. The jump
amplitudes are described by the pdf
$h(x,u)dx=\mbox{Prob}\{x<A_k(u)<x+dx\}$ and the jump times $T_k(u)$ follow
a Poisson distribution of parameter $\lambda(u)$. Define 
$$
p(x_1,t_1;x_2,t_2;u)dx_1dx_2=\mbox{Prob}\{x_1<Y(u,t_1)<x_1+dx_1;
x_2<Y(u,t_2)<x_2+dx_2\}
$$
to be the joint pdf of the process with 
$t_2\geq t_1$. This pdf can be written as
\begin{equation}
p(x_1,t_1;x_2,t_2;u)=\sum_{n_1=0}^{\infty}\sum_{n_2=0}^{\infty}
p(x_1,t_1;x_2,t_2;u|n_1,n_2)P(n_1,t_1;n_2,t_2;u),
\label{a2}
\end{equation}
where $p(x_1,t_1;x_2,t_2;u|n_1,n_2)$ is the conditional pdf assuming that
exactly $n_1$ pulses have occurred at time $t_1$ and $n_2$ pulses at time
$t_2$.
$P(n_1,t_1;n_2,t_2;u)$ is the joint probability for the occurrence of such
pulses. Since $t_2\geq t_1$ then $n_2\geq n_1$ and  
\begin{equation}
P(n_1,t_1;n_2,t_2;u)=
\cases{P(n_2-n_1;t_2-t_1;u)P(n_1,t_1;u),&if $n_1\leq n_2$\cr
                 0, &otherwise,\cr}
\label{a3}
\end{equation}
where 
\begin{equation}
P(m,\tau;u)=\frac{[\lambda(u)\tau]^m}{m!}e^{-\lambda(u)\tau}
\label{a4}
\end{equation}
is the Poisson distribution. Substituting Eq.~(\ref{a3}) into
Eq.~(\ref{a2}) and defining $t_1=t$, $t_2=t+\Delta t$, $n_1=n$ and
$n_2-n_1=m$, we obtain
$$
p(x_1,t;x_2,t+\Delta t;u)=\sum_{n=0}^{\infty}\sum_{m=0}^{\infty}
p(x_1,t;x_2,t+\Delta t;u|n,n+m)P(m,\Delta t;u)P(n,t;u),
$$
and the characteristic function reads
\begin{equation}
\tilde{p}(\omega_1,t;\omega_2,t+\Delta
t;u)=\sum_{n=0}^{\infty}\sum_{m=0}^{\infty}
\tilde{p}(\omega_1,t;\omega_2,t+\Delta t;u|n,n+m)
P(m,\Delta t;u)P(n,t;u).
\label{a5}
\end{equation}
Note that $\tilde{p}(\omega_1,t;\omega_2,t+\Delta t;u|n,n+m)$ is the joint
characteristic function of the truncated process
$$
Y_n(u,t)=\sum_{k=1}^{n}A_k(u)\phi\left[t-T_k(u); u\right].
$$
Hence
\begin{eqnarray*}
\tilde{p}(\omega_1,t;\omega_2,t+\Delta t;u|n,n+m)
=\Biggl\langle
\exp\Bigl[i\omega_2\sum_{k=1}^{n}A_k(u)
\phi\bigl(t+\Delta t&-&T_k(u);u\bigr)
+i\omega_1\sum_{l=1}^{n}A_l(u)\phi\bigl(t-T_l(u);u\bigr)\nonumber \\ 
&+&
i\omega_2\sum_{k=n+1}^{n+m}A_k(u)
\phi\bigl(t+\Delta t-T_k(u);u\bigr)\Bigr]\Biggr\rangle. 
\end{eqnarray*}
Taking into account that $A_k(u)$ and $T_k(u)$ are independent and
identically distributed random variables, we have
\begin{eqnarray*}
\tilde{p}(\omega_1,t;\omega_2,t+\Delta t;u|n,n+m)=&&
\Bigl[\bigl\langle\exp\{i\omega_2 A_k(u)\phi(t+\Delta
t-T_k(u);u)+i\omega_1
A_k(u)\phi(t-T_k(u);u)\}\bigr\rangle\Bigr]^{n}\nonumber \\
&&\times\Bigl[\bigl\langle\exp\{i\omega_2 A_k(u)\phi(t+\Delta
t-T_k(u);u)\}\bigr\rangle\Bigr]^{m},
\end{eqnarray*}
and, since the random times $T_k(u)$ are Poissonian,
\begin{eqnarray*}
\langle\exp\{i\omega_2 A_k(u)\phi(t+\Delta t-T_k(u);u)&+&i\omega_1
A_k(u)\phi(t-T_k(u);u)\}\rangle \nonumber\\
&=&\int_{-\infty}^{\infty}h(a,u)da\int_0^t\frac{dt'}{t}
\exp\{ia[\omega_2\phi(t+\Delta t-t';u)+\omega_1\phi(t-t';u)]\}\nonumber\\
&=&\frac{1}{t}\int_0^t\tilde{h}
\bigl[\omega_2\phi(t+\Delta t-t';u)+\omega_1\phi(t-t';u); u\bigr]dt',
\end{eqnarray*}
where $\tilde{h}(\omega;u)$ is the Fourier transform of the jump pdf
$h(x;u)$. Analogously
$$
\langle\exp\{i\omega_2 A_k(u)\phi(t+\Delta t-T_k(u);u)\}\rangle =
\frac{1}{\Delta t}\int_t^{t+\Delta t}\tilde{h}
\bigl[\omega_2\phi(t+\Delta t-t';u);u\bigr]dt'.
$$
Therefore,
\begin{equation}
\tilde{p}(\omega_1,t;\omega_2,t+\Delta t;u|n,n+m)=
\left[\frac{1}{t}F(\omega_2,t+\Delta t;\omega_1,t;u)\right]^{n}
\left[\frac{1}{\Delta t}G(\omega_2,t+\Delta t;t;u)\right]^{m},
\label{a6}
\end{equation}
where
\begin{equation}
F(\omega_2,t+\Delta
t;\omega_1,t;u)\equiv\int_0^t\tilde{h}\bigl[\omega_2\phi(t+\Delta t-t';u)+
\omega_1\phi(t-t';u);u\bigr]dt',
\label{a7}
\end{equation}
and
\begin{equation}
G(\omega_2,t+\Delta t;t;u)\equiv\int_t^{t+\Delta t}
\tilde{h}\bigl[\omega_2\phi(t+\Delta t-t';u);u\bigr]dt'.
\label{a8}
\end{equation}
Substituting Eq.~(\ref{a6}) into Eq.~(\ref{a5}) yields
$$
\tilde{p}(\omega_1,t;\omega_2,t+\Delta t;u)=
\left\{\sum_{n=0}^\infty\left[\frac{1}{t}F(\omega_1,t;\omega_2,t+\Delta
t;u)\right]^{n}P(n,t;u)\right\}
\left\{\sum_{m=0}^\infty\left[\frac{1}{\Delta t}
G(\omega_2,t+\Delta t;t;u)\right]^{m}P(m,\Delta t;u)\right\}.
$$
Introducing Eq.~(\ref{a4}) into this and performing the resulting sums we
get
$$
\tilde{p}(\omega_1,t;\omega_2,t+\Delta t;u)=
\exp\left\{-\lambda(u)\left[t-F(\omega_2,t+\Delta t;\omega_1,t;u)\right]-
\lambda(u)\left[\Delta t-G(\omega_2,t+\Delta
t;\omega_1,t;u)\right]\right\}.
$$
Finally
\begin{eqnarray}
\tilde{p}(\omega_1,t;\omega_2,t+\Delta t;u)=\exp\Biggl\{\lambda(u)
\int_0^{t}dt'\Bigl[\tilde{h}\bigl[\omega_1\phi(t',u)&+&
\omega_2\phi(t'+\Delta t,u)\bigr]-1\Bigr]\nonumber \\
&+&
\lambda(u)\int_0^{\Delta t}dt'\Bigl[
\tilde{h}\bigl[\omega_2\phi(t',u)\bigr]-1\Bigr]\Biggl\},
\label{a9}
\end{eqnarray}
which agrees with Eq.~(\ref{5a}). If in~(\ref{a9}) we set
$\omega_1=\omega$ and $\omega_2=0$, we obtain the one time characteristic
function~(\ref{5}):
\begin{equation}
\tilde{p}(\omega,t;u)=
\exp\left\{\lambda(u)\int_0^{t}dt'
\left[\tilde{h}\left[\omega\phi(t',u)\right]-1\right]\right\}.
\label{a10}
\end{equation}

\section{Cumulants for a step-like function}

We will derive closed expressions for the cumulants~(\ref{cumulant2}):
\begin{equation}
\langle\langle X^{2n}(t)\rangle\rangle=
\frac{(-1)^{n}\tilde{h}^{(2n)}(0)}{2n}(bt)^{2n/\alpha}
\int_0^\infty\frac{\phi^{2n}(x)}{x^{2n/\alpha}}dx.
\label{b1}
\end{equation}
when $\phi(x)$ is the step-like function~(\ref{24b}):
\begin{equation}
\phi(t)=\cases{1-e^{-kt}, &if $t>0$\cr
                 0, &otherwise.\cr}
\label{b2}
\end{equation}
The substitution of Eq.~(\ref{b2}) into Eq.~(\ref{b1}) leads us to
evaluate the following integral
\begin{eqnarray*}
I_{2n}&=&\int_0^\infty\frac{(1-e^{-kx})^{2n}}{x^{2n/\alpha}}dx\\
&=&\int_0^\infty\frac{dx}{x^{2n/\alpha}}
\biggl[kx\int_0^1e^{-kxu}du\biggr]^{2n}\\
&=&k^{2n}\int_0^1du_1\cdots\int_0^1du_{2n}\int_0^{\infty}
x^{2n(1-1/\alpha)}
\exp\Biggl\{-kx\Biggl(\sum_{i=1}^{2n}u_i\Biggr)\Biggr\}dx.
\end{eqnarray*}
Define the new integration variable 
$$
\xi=kx\Biggl(\sum_{i=1}^{2n}u_i\Biggr),
$$
then
$$
I_{2n}=k^{-1+2n/\alpha}\int_0^1du_1\cdots\int_0^1du_{2n}
\left(\sum_{i=1}^{2n}u_i\right)^{-1-2n(1-1/\alpha)}
\int_0^{\infty}e^{-\xi}\xi^{2n(1-1/\alpha)}d\xi.
$$
But 
$$
\int_0^{\infty}e^{-\xi}\xi^{2n(1-1/\alpha)}d\xi=\Gamma(1+2n-2n/\alpha).
$$
Defining the numbers 
\begin{equation}
A_{2n}\equiv\int_0^1du_1\cdots\int_0^1du_{2n}
\left(\sum_{i=1}^{2n}u_i\right)^{-1-2n(1-1/\alpha)},
\label{b3}
\end{equation}
we finally have 
\begin{equation}
\langle\langle X^{2n}(t)\rangle\rangle=
\frac{(-1)^{-n}\tilde{h}^{(2n)}(0)}{2n}A_{2n}
k^{-1+2n/\alpha}\Gamma(1+2n-2n/\alpha)(bt)^{2n/\alpha},
\label{b4}
\end{equation}
which is Eq.~(\ref{cumulant3}).

\end{document}